\documentclass[aps,prd,groupedaddress,nofootinbib,preprint]{revtex4}

\usepackage{ulem}
\usepackage{color}
\usepackage{graphicx}
\usepackage{dcolumn}
\usepackage{bm}

\begin{document}

\title{Constraining Absolute Neutrino Masses via Detection of \\ Galactic Supernova Neutrinos at JUNO}

\author{Jia-Shu Lu}

\email{lujiashu@ihep.ac.cn}

\affiliation{Institute of High Energy Physics, Chinese Academy of Sciences, Beijing 100049, China}

\author{Jun Cao}

\email{caoj@ihep.ac.cn}

\affiliation{Institute of High Energy Physics, Chinese Academy of Sciences, Beijing 100049, China}

\author{Yu-Feng Li}

\email{liyufeng@ihep.ac.cn}

\affiliation{Institute of High Energy Physics, Chinese Academy of Sciences, Beijing 100049, China}

\author{Shun Zhou}

\email{zhoush@ihep.ac.cn}

\affiliation{Institute of High Energy Physics, Chinese Academy of Sciences, Beijing 100049, China}

%\affiliation{Center for High Energy Physics, Peking University, Beijing %100871, China}

\begin{abstract}
A high-statistics measurement of the neutrinos from a galactic core-collapse supernova is extremely important for understanding the explosion mechanism, and studying the intrinsic properties of neutrinos themselves. In this paper, we explore the possibility to constrain the absolute scale of neutrino masses $m^{}_\nu$ via the detection of galactic supernova neutrinos at the Jiangmen Underground Neutrino Observatory (JUNO) with a 20 kiloton liquid-scintillator detector. In assumption of a nearly-degenerate neutrino mass spectrum and a normal mass ordering, the upper bound on the absolute neutrino mass is found to be $m^{}_\nu < (0.83 \pm 0.24)~{\rm eV}$ at the 95\% confidence level for a typical galactic supernova at a distance of 10 kpc, where the mean value and standard deviation are shown to account for statistical fluctuations. For comparison, we find that the bound in the Super-Kamiokande experiment is $m^{}_\nu < (0.94 \pm 0.28)~{\rm eV}$ at the same confidence level. However, the upper bound will be relaxed when the model parameters characterizing the time structure of supernova neutrino fluxes are not exactly known, and when the neutrino mass ordering is inverted.
\end{abstract}

\maketitle

\section{Introduction}

The neutrino burst from Supernova (SN) 1987A in the Large Magellanic Cloud  was clearly recorded in the Kamiokande-II~\cite{Hirata:1987hu}, Irvine-Michigan-Brookhaven (IMB)~\cite{Bionta:1987qt}, and Baksan~\cite{Alekseev:1988gp} experiments. Although only twenty-four neutrino events in total were observed, we have acquired very useful information about the explosion mechanism of core-collapse SNe and the intrinsic properties of neutrinos themselves~\cite{Raffelt:1996bk}. As pointed out by Zatsepin long time ago, neutrinos from SNe can be used to constrain the absolute neutrino masses through their delayed flight time~\cite{Zatsepin:1968kt}. In fact, the observed neutrino events associated with SN 1987A have been reanalyzed by Loredo and Lamb in great detail~\cite{Loredo:2001rx}, and they have obtained a restrictive limit $m^{}_\nu < 5.7~{\rm eV}$ at $95\%$ confidence level (CL)\footnote{For simplicity, we hereafter assume the neutrino mass spectrum to be quasi-degenerate, i.e., $m^{}_1 \approx m^{}_2 \approx m^{}_3 \equiv m^{}_\nu$. Hence all the experimental bounds on absolute neutrino masses can be translated to those on the neutrino mass scale $m^{}_\nu$. In particular, this assumption is well justified for SN neutrinos, since current experiments are only sensitive to neutrino masses lying in the quasi-degenerate region.}, significantly improving the even earlier analyses in the literature~\cite{Adams:1988im, Abbott:1987bm, Spergel:1987ex, Burrows:1988, Loredo:1988mk, Kernan:1994kt}.

Recently, the upper limit on neutrino masses from the neutrino detection of a galactic SN has been derived in Ref.~\cite{Pagliaroli:2010ik} for a large water Cherenkov detector, such as the Super-Kamiokande experiment (SK)~\cite{Fukuda:2002uc}. It has been shown that $m^{}_\nu < 0.8~{\rm eV}$ at the $95\%$ CL can be achieved for a typical galactic SN at a distance of $10~{\rm kpc}$. An upper bound in the sub-eV region has also been claimed in Refs.~\cite{Nardi:2003pr} and \cite{Nardi:2004zg} for future large water Cherenkov and scintillator detectors (e.g., Hyper-Kamiokande~\cite{Nakamura:2003hk} and LENA~\cite{Wurm:2011zn}). In this paper, we investigate the sensitivity of JUNO to absolute neutrino masses via the detection of galactic SN neutrinos. The motivation for such an investigation is as follows:
\begin{itemize}
\item For a galactic SN neutrino burst, the JUNO detector is expected to collect about $10^4$ events, mainly in the inverse beta decay channel $\overline{\nu}^{}_e + p \to e^+ + n$ (IBD). The total number of IBD events is comparable to that in SK. Furthermore, the JUNO detector has a lower energy threshold and a better energy resolution, which are crucial advantages in measuring the time-delay effects of low-energy SN neutrinos and should lead to an improvement of the neutrino mass bound.

\item An independent probe of absolute neutrino masses in the sub-eV region is necessary and desirable. Thus far, the most stringent constraint on the effective neutrino mass in tritium beta decays is $m^{}_\nu < 2.1~{\rm eV}$ at $95\%$ CL from the Troitsk Collaboration~\cite{Aseev:2011dq}. The future KATRIN~\cite{Osipowicz:2001sq} and Project 8~\cite{Formaggio:2011ba} experiments are planning to improve the present limit by one order of magnitude, reaching the sub-eV level. Useful information on neutrino masses can also be obtained from the detection of neutrinoless double-beta decays, if neutrinos are Majorana particles, i.e., neutrinos are their own antiparticles. Given the current bound on the effective neutrino mass $\langle m \rangle^{}_{\beta \beta} \equiv \left|U^2_{e1} m^{}_1 + U^2_{e2} m^{}_2 + U^2_{e3} m^{}_3\right| \lesssim 0.4~{\rm eV}$~\cite{Rodejohann:2012}, where $U^{}_{e i}$ (for $i = 1, 2, 3$) stand for three elements in the first row of neutrino mixing matrix $U$, one can derive $m^{}_\nu \lesssim (0.4 \cdots 1.5)~{\rm eV}$, depending on the neutrino mass hierarchy, neutrino mixing angles and two unknown Majorana CP-violating phases.

\item Cosmological observations give rise to a very tight bound on the sum of three neutrino masses, namely $\Sigma \equiv m^{}_1 + m^{}_2 + m^{}_3 < 0.23~{\rm eV}$ at the $95\%$ CL~\cite{Ade:2013zuv}. However, it should be noted that the cosmological bound can be quite different if the systematic uncertainties for some data sets are taken into account or if an extension of the minimal standard model of cosmology is considered~\cite{Giusarma:2013pmn}. Therefore, it is interesting to investigate whether a high-statistics observation of galactic SN neutrinos could provide an independent and competitive bound on $m^{}_\nu$.
\end{itemize}

Although following closely the general approach proposed in Ref.~\cite{Pagliaroli:2010ik}, our analysis differs from previous works in several aspects. First, we compare the mass bound from JUNO with that from SK, and point out that JUNO with a lower energy threshold and a higher energy resolution can indeed improve the bound. Due to the limited statistics in the low-energy and early-time region, however, such an improvement is moderate. Second, we have simulated a large number of experiments to study the statistical uncertainties of neutrino mass bound. The upper bound at the $95\%$ CL for JUNO turns out be $m^{}_\nu < (0.83\pm 0.24)~{\rm eV}$, while that for SK $m^{}_\nu < (0.94\pm 0.28)~{\rm eV}$.
Third, we point out that both the starting time $t^{}_{\rm s}$ and the rising-time interval $\tau^{}_{\rm r}$ for neutrino emission have crucial impact on the neutrino mass bound. If these two SN model parameters are set to be free, the upper bound will be relaxed to $m^{}_\nu < (1.12\pm 0.33)~{\rm eV}$ for JUNO, and $m^{}_\nu < (1.49 \pm 0.42)~{\rm eV}$ for SK.

The remaining part of this work is organized as follows. In Sec. II, we briefly review a simple model of the SN $\overline{\nu}^{}_e$ flux, and discuss the general approach to constrain absolute neutrino mass scale by observing the time-delay effects of SN neutrinos. Sec. III is devoted to our simulation results for JUNO, and a comparison is made with SK. Furthermore, we investigate the impact of model parameters on the mass bound, and consider the statistical uncertainty by simulating a large number of experiments. The systematic uncertainties induced by different SN models are also briefly discussed. Finally we conclude and summarize our main results in Sec. IV.

\section{SN Neutrinos and Mass Bound}

Since neutrinos are massive, their flight time from the SN to the detector at the Earth will be delayed, compared to that of massless particles~\cite{Zatsepin:1968kt}. For the neutrinos from a galactic core-collapse SN, the time delay can be written as
\begin{equation}
\Delta t(m^{}_\nu, E^{}_\nu) \simeq 5.14~{\rm ms}~\left( \frac{m^{}_\nu}{\rm eV} \right)^2 \left( \frac{10~{\rm MeV}}{E^{}_\nu} \right)^2 \frac{D}{10~{\rm kpc}} \; ,
%     (1)
\end{equation}
where $E^{}_\nu$ is the neutrino energy, and $D$ is the distance between SN and detector. Thus, a time delay at the millisecond level is expected for the neutrinos of eV masses from a typical galactic SN at $D = 10~{\rm kpc}$. If the neutrino energies and arrival time can be precisely measured, the shifts of neutrino events in the time distribution will signify nonzero neutrino masses. However, the average energy and luminosity of SN neutrinos evolve in time, which complicates the extraction of neutrino mass information from experimental observations. In order to derive a mass bound, one has to model the time evolution of neutrino energies and fluxes, and take into account neutrino flavor conversions in the propagation from the SN core to the detector.

\subsection{Parameterized Neutrino Flux}

In a core-collapse SN, neutrinos and antineutrinos of three different flavors can be produced~\cite{Janka:2006fh}. Along with the development of SN explosion, neutrino emission can be described in three distinct stages. (1) Neutronization $\nu^{}_e$ burst: Just after bounce, a shock wave forms and dissociates heavy nuclei, so the electron capture on free protons $e^- + p \to \nu^{}_e + n$ gives rise to a burst of $\nu^{}_e$. (2) Accretion phase: The prompt shock wave looses its energy in disintegrating heavy nuclei and eventually turns into an accreting shock, where the hot $e^+ e^-$ plasma will generate intense $\overline{\nu}^{}_e$ and $\nu^{}_e$ luminosities via charged-current interactions with free nucleons. (3) Cooling phase: After the explosion, a proto-neutron star forms in the center and cools down by emitting neutrinos and antineutrinos of all flavors.

As the cross section of IBD is much larger than those of other channels for SN neutrinos of typical energies, we only concentrate on the $\overline{\nu}^{}_e$ flux $\Phi^0_{\overline{\nu}^{}_e}$ in the accretion and cooling phases. In Ref.~\cite{Pagliaroli:2008ur}, a simple parametrization of $\Phi^0_{\overline{\nu}^{}_e}$ is presented to capture essential physics of SN neutrino production and the main features of numerical simulations. In the cooling phase, the $\overline{\nu}^{}_e$ flux $\Phi^0_{\rm c}$ is parameterized by three model parameters: the initial temperature $T^{}_{\rm c}$, the radius of neutrino sphere $R^{}_{\rm c}$, and the cooling time scale $\tau^{}_{\rm c}$. In the accretion phase, one has to model the time evolution of neutron number and positron temperature in order to figure out the $\overline{\nu}^{}_e$ flux $\Phi^0_{\rm a}$. This can be done by introducing the accretion time scale $\tau^{}_{\rm a}$, and requiring the resultant neutrino energy and luminosity to approximately follow numerical simulations. In addition, the initial number of neutrons depends on an initial accreting mass $M^{}_{\rm a}$, and a thermal energy spectrum of positrons depending on an initial temperature $T^{}_{\rm a}$ is assumed. Putting all together, the total flux is~\cite{Pagliaroli:2008ur, Pagliaroli:2009qy}
\begin{equation}
\Phi^0_{\overline{\nu}^{}_e}(t, E^{}_\nu) = f^{}_{\rm r}(t) \Phi^0_{\rm a}(t, E^{}_\nu) + \left[1 - j^{}_k(t)\right] \Phi^0_{\rm c}(t, E^{}_\nu) \; ,
%     (2)
\end{equation}
where $f^{}_{\rm r}(t) = 1 - \exp(-t/\tau^{}_{\rm r})$ with the rising time scale $\tau^{}_{\rm r}$ further introduces an early-time fine structure, and $j^{}_k(t) = \exp[- (t/\tau^{}_{\rm a})^k]$ (with $k$ being an integer) is the time function interpolating the accretion and cooling phases of neutrino emission. In our calculations, $k = 2$ will be chosen, and the analytical expressions of parameterized fluxes $\Phi^0_{\rm a}(t, E^{}_\nu)$ and $\Phi^0_{\rm c}(t, E^{}_\nu)$, which can be found in Ref.~\cite{Pagliaroli:2008ur}, will be implemented.

\subsection{Neutrino Flavor Conversions}

When propagating from the SN core to the envelope, neutrinos experience three different stages of flavor oscillations. First, the matter density in the SN core is so high that the coherent flavor conversions will be interrupted by the frequent scattering of neutrinos on matter particles. Therefore, the lepton flavors are indeed conserved in the dense core~\cite{Hannestad:1999zy}. Even for the heavy-lepton flavors $\nu^{}_\mu$ and $\nu^{}_\tau$, the one-loop corrections to neutrino refractive index are significant enough to enhance matter effects, suppressing the flavor oscillations. Second, from the neutrino sphere to the radius of several hundred kilometers, neutrino number densities are even higher than or comparable to the electron number density of matter, and thus the neutrino-neutrino refraction may lead to an instability in the flavor space. As a consequence, collective neutrino oscillations may take place, and cause remarkable changes in the neutrino energy spectra~\cite{Duan:2005cp,Duan:2006an}. The flavor instability induced by neutrino self-interaction is currently an unresolved problem~\cite{Duan:2010bg}. It has been argued that the high matter density during the accretion phase could completely suppress collective neutrino oscillations~\cite{EstebanPretel:2008ni, Chakraborty:2011nf, Sarikas:2011am}. More recently, the axial symmetry usually assumed in the previous studies of SN neutrino oscillations has been found to be not satisfied by the solutions to the equations of neutrino flavor evolution~\cite{Raffelt:2013rqa, Raffelt:2013isa, Mirizzi:2013rla, Mirizzi:2013wda, Duan:2013kba}, and the flavor instability needs more dedicated investigations. As for the time-delay effects induced by neutrino masses, only the low-energy neutrinos from the early accretion phase are quite relevant. In this case, we expect that the dense matter will highly suppress the growth of flavor instability, and thus the collective neutrino oscillations do not take place at all. However, it should be noticed that a dedicated study should be carried out to clarify whether or not collective oscillations actually occur and significantly change neutrino flavors and energy spectra. In the present work, we temporarily put aside the collective neutrino oscillations in order to avoid further complications. Finally, the resonant flavor conversions corresponding to neutrino mass-squared differences $|\Delta m^2_{31}| \equiv |m^2_3 - m^2_1| = 2.4\times 10^{-3}~{\rm eV}^2$ and $\Delta m^2_{21} \equiv m^2_2 - m^2_1 = 7.5\times 10^{-5}~{\rm eV}^2$~\cite{PDG2014} will occur in the SN envelope, where the matter density becomes suitable for the Mikheyev-Smirnov-Wolfenstein effects to be crucially important~\cite{Mikheev:1986gs,Wolfenstein:1977ue}.

The Mikheyev-Smirnov-Wolfenstein effects on neutrino flavor oscillations in the SN envelope have been examined in detail in Ref.~\cite{Dighe:1999bi}. Taking account of a relatively large $\theta^{}_{13}$ (e.g., $\sin^2 \theta^{}_{13} \approx 0.024$), which has been precisely measured in Daya Bay and other reactor experiments~\cite{An:2012eh,An:2013uza,DC,RENO}, one can find that the transition associated with the $\Delta m^2_{31}$ resonance is perfectly adiabatic. Therefore, if the neutrino mass ordering is normal, i.e., $\Delta m^2_{31} > 0$ or $m^{}_1 < m^{}_2 < m^{}_3$, the flux of $\overline{\nu}^{}_e$ on the surface of SN is given by
\begin{equation}
\Phi^{}_{\overline{\nu}^{}_e} = \cos^2 \theta^{}_{12} \Phi^0_{\overline{\nu}^{}_e} + \sin^2 \theta^{}_{12} \Phi^0_{\nu^{}_x}
\end{equation}
with $\sin^2 \theta^{}_{12} \approx 0.302$, where $\Phi^0_{\overline{\nu}^{}_e}$ and $\Phi^0_{\nu^{}_x}$ denote the fluxes of $\overline{\nu}^{}_e$ and $\nu^{}_x$ in the case of no flavor oscillations, respectively. If the neutrino mass ordering is inverted, i.e., $\Delta m^2_{31} < 0$ or equivalently $m^{}_3 < m^{}_1 < m^{}_2$, we have $\Phi^{}_{\overline{\nu}^{}_e} = \Phi^0_{\nu^{}_x}$. Note that $\Phi^0_{\overline{\nu}^{}_e}$ receives the contributions from both accretion and cooling phases, as indicated by Eq. (2), while $\Phi^0_{\nu^{}_x}$ is assumed to be vanishing in the accretion phase~\cite{Pagliaroli:2008ur}. Thus, the total flux of $\overline{\nu}^{}_e$ in the case of inverted neutrino mass ordering is much lower than that in the case of normal mass ordering, leading to a reduced number of IBD events and less restrictive bound on neutrino masses. In the following discussions, we shall focus on the normal neutrino mass ordering, but it should be noticed that the derived bound is the most optimistic one.

\subsection{Numerical Method}

Given the flux $\Phi^{}_{\overline{\nu}^{}_e}(t, E^{}_\nu)$, one can obtain the event rate $R(t, E^{}_e)$ by convolving it with the IBD cross section $\sigma^{}_{\rm IBD}(E^{}_\nu)$, which can be found in Ref.~\cite{Strumia:2003zx}. In the IBD reaction, the positron energy is approximately given by $E^{}_e \approx E^{}_\nu - \Delta$ with $\Delta = m^{}_n - m^{}_p \approx 1.293~{\rm MeV}$ being the neutron-proton mass difference. More explicitly, we have
\begin{equation}
R(t, E^{}_e) = N^{}_p \Phi^{}_{\overline{\nu}^{}_e}(t, E^{}_\nu) \sigma^{}_{\rm IBD}(E^{}_\nu) \eta(E^{}_e) \; ,
\end{equation}
where $N^{}_p$ is the number of target protons in the detector, and the detection efficiency factor $\eta(E^{}_e) = 1$ is taken for JUNO, while $\eta(E^{}_e) = 0.98$ for SK. Note that the angular distribution of the final-state positron is almost isotropic, since the kinetic energy of positron is much lower than the nucleon mass. The visible energy is $E^\prime = E^{}_e + m^{}_e \approx E^{}_\nu - 0.8~{\rm MeV}$, which will be observed as $E$ in the detector due to a finite energy resolution $\delta E$. For simplicity, the Gaussian distribution $G(E^\prime, E; \delta E)$ for the energy smearing is assumed.

Our strategy to generate neutrino events and derive the bound on absolute neutrino masses is as follows. First, the event rate $R(t, E^{}_e)$ is used as a target distribution function to randomly produce $N$ neutrino events, represented by $(t^{}_i, E^{}_i)$ for $i = 1, 2, \cdots, N$, where the observed energy $E^{}_i$ has further been generated according to the Gaussian smearing function $G(E^{}_e + m^{}_e, E; \delta E)$ for each event. Second, we implement the neutrino flux model with six astrophysical parameters (i.e., $\tau^{}_{\rm a}$, $T^{}_{\rm a}$ and $M^{}_{\rm a}$ for the accretion phase, $\tau^{}_{\rm c}$, $T^{}_{\rm c}$ and $R^{}_{\rm c}$ for the cooling phase), the rising time $\tau^{}_{\rm r}$ for the early time structure, the absolute starting time $t^{}_{\rm s}$ for neutrino emission and the neutrino mass $m^{}_\nu$ to fit the previously generated neutrino events. In order to make use of the time information of every single event, we define the following likelihood function~\cite{Pagliaroli:2008ur}
\begin{equation}
{\cal L} = e^{-\int^{T}_0 R(t) {\rm d}t} \prod^{N}_{i = 1} \int^{\infty}_{E^{}_{\rm th}} R(t^\prime_i, E^{}_e) G(E^{}_e + m^{}_e, E^{}_i; \delta E^{}_i) {\rm d}E^{}_e \; ,
\end{equation}
where $t^\prime_i = t^{}_i - \Delta t(m^{}_\nu, E^i_\nu) - t^{}_{\rm s}$ stands for the real time when the corresponding neutrino is emitted. Note that a common time needed for massless particles to travel from the SN to the detector has been subtracted from the measured time, and the neutrino energy is constructed from the detected energy via $E^i_\nu = E^{}_i + 0.8~{\rm MeV}$. Then, the theoretical model parameters can be estimated by $\chi^2 = -2 \ln {\cal L}$, and the test statistic is defined as $\Delta \chi^2 \equiv \chi^2(m^{}_\nu) - \chi^2_{\rm min}$. Third, to take into account the statistical fluctuation, we simulate a large number of experiments, for which the total number of events is chosen according to the Poisson distribution with an expectation value that is calculated by integrating the rate $R(t, E^{}_e)$ over the time duration $T = 30~{\rm s}$ and the positron energy above its threshold.

In the following, we shall perform the numerical simulations for JUNO and SK in order to derive the SN bound on absolute neutrino masses. Furthermore, the impact of astrophysical model parameters on the mass bound and the statistical uncertainties are studied.

\section{Results and Discussions}

For JUNO~\cite{DYB2e,DYB2t,Li:2013zyd}, a fiducial mass of 20 kiloton liquid scintillator with a proton fraction of $12\%$ is adopted, and the energy resolution is assumed to be $\delta E/E = 0.03 \sqrt{{\rm MeV}/E}$.
Using the SN neutrino flux model in Sec. II, we obtain the total number of IBD events $N^{}_{\rm a} = 4516$ in the JUNO detector. In each simulation, the actual number of IBD events is determined according to the Poisson distribution with $N^{}_{\rm a}$ as the expectation value.

%%%%%%%%%%%%%%%%%%%%%%%%%% Fig. 1 %%%%%%%%%%%%%%%%%%%%%%%%%%%%%%%%%%%%%%%%%%
\begin{figure}
    \centering
    \includegraphics[width=0.6\textwidth]{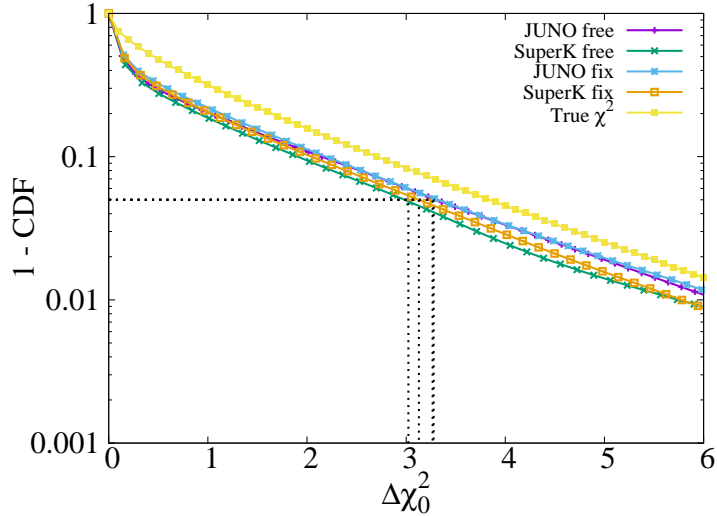}
    \caption{Distribution of the test statistic $\Delta \chi^2_0$ for JUNO and SK. The cumulative distribution function (CDF) of $\Delta \chi^2_0$ has been calculated for four different cases, where the situations with other parameters being free or fixed are explicitly indicated (see the text for details). For comparison, the true $\chi^2$ distribution with one degree of freedom is also shown. The values of $\Delta \chi^2_0$ corresponding to the $95\%$ CL (i.e., the horizontal dashed line) are 3.12 (SK fixed), 3.28 (JUNO fixed), 3.02 (SK free), and 3.27 (JUNO free), whereas $\Delta \chi^2_0 = 3.84$ for the $\chi^2$ distribution.}
\end{figure}
%%%%%%%%%%%%%%%%%%%%%%%%%%%%%%%%%%%%%%%%%%%%%%%%%%%%%%%%%%%%%%%%%%%%%%%%%%%%

In order to generate the artificial data, we first set the astrophysical model parameters to their best-fit values derived from the neutrino data of SN 1987A~\cite{Pagliaroli:2008ur}: $R^{}_{\rm c} = 16.0~{\rm km}$, $T^{}_{\rm c} = 4.6~{\rm MeV}$, $\tau^{}_{\rm c} = 4.7~{\rm s}$, $M^{}_{\rm a} = 0.22~M^{}_\odot$, $T^{}_{\rm a} = 2.4~{\rm MeV}$ and $\tau^{}_{\rm a} = 0.55~{\rm s}$. In addition, the rising time is chosen to be $\tau^{}_{\rm s} = 50~{\rm ms}$, while the starting time $t^{}_{\rm s}$ and the neutrino mass $m^{}_\nu$ are set to zero. Note that the distance is fixed as $D = 10~{\rm kpc}$ in all simulations. Then, the artificial data are used to construct the likelihood function in Eq.~(5). To examine whether the observations are consistent with the hypothesis of $m^{}_\nu = 0$, we define the test statistic as $\Delta \chi^2_0 \equiv \chi^2(m^{}_\nu = 0) - \chi^2_{\rm min}$, which is equivalently the logarithm of the likelihood ratio $\Delta \chi^2_0 = -2 \ln {\cal L}(m^{}_\nu = 0)/{\cal L}^{}_{\rm max}$. Given each artificial data set, the $\chi^2$ function is minimized to calculate $\Delta \chi^2_0$. Then, a large number of simulations are performed to obtain the distribution function $f^{}_0(\Delta \chi^2_0)$ of the test statistic~\cite{Schwetz:2006md}. The hypothesis will be excluded at the $100(1 - \alpha)\%$ CL, if the test statistic is larger than $\lambda(\alpha)$. The latter is fixed by requiring the integration of $f^{}_0(\Delta \chi^2_0)$ over the range $\Delta \chi^2_0 > \lambda(\alpha)$ to be equal to $\alpha$.

In Fig.~1, the cumulative distribution function of $\Delta \chi^2_0$ has been shown for JUNO in two different cases. In the first case, the parameters $\tau^{}_{\rm r}$ and $t^{}_{\rm s}$ are fixed at their true values. In the second one, they are set to be free and the condition $\Delta \chi^2_0(\tau^{}_{\rm r}, t^{}_{\rm s}) > \lambda(\alpha)$ is satisfied for all possible values of $\tau^{}_{\rm r}$ and $t^{}_{\rm s}$.

For SK~\cite{Fukuda:2002uc}, we assume that the fiducial mass is 22.5 kt and the proton fraction is $11\%$. In addition, the threshold of visible energy is taken as $6.5~{\rm MeV}$, while the energy resolution $\delta E/E = 0.023 + 0.41\sqrt{{\rm MeV}/E}$. Using the same SN neutrino fluxes, one can get the total number of IBD events is $N^{}_{\rm a} = 4451$, which is very close to the number at JUNO. The distribution of $\Delta \chi^2_0$ for SK is also calculated by following the above strategy and the result is shown in Fig.~1, where two cases with fixed and free values of $\tau^{}_{\rm r}$ and $t^{}_{\rm s}$ are considered.

For comparison, the true $\chi^2$ distribution for one degree of freedom is depicted in Fig.~1. To draw a neutrino mass bound at the $95\%$ CL, one usually assumes the $\chi^2$ distribution and requires $\Delta \chi^2_0 > 3.84$. As one can observe from Fig.~1, the real distribution of $\Delta \chi^2_0$ deviates insignificantly from the $\chi^2$ one, for both JUNO and SK. The critical values of $\Delta \chi^2_0$, above which the hypothesis of $m^{}_\nu = 0$ will be excluded at the $95\%$ CL, have been indicated in Fig.~1 and summarized in the caption. Therefore, the neutrino mass bound at the $95\%$ CL in the present work should be interpreted as $\Delta \chi^2_0\simeq 3.28$ (or $3.27$) for JUNO, and as $\Delta \chi^2_0\simeq 3.12$ (or $3.02$) for SK, for the statistical method with fixed (or free) astrophysical time parameters. For a specific simulation, we calculate $\Delta \chi^2 = \chi^2(m^{}_\nu) - \chi^2_{\rm min}$ and set the $95\%$ CL upper bound by requiring $\Delta \chi^2$ to follow the corresponding distribution in Fig.~1.

\subsection{Impact of Model Parameters}
To examine the impact of model parameters on the neutrino mass bound at JUNO, we study a single experiment and analyze the parameter fit to the artificial data in more detail. In Fig.~2, we randomly take one simulation with $N = 4499$ IBD events, and calculate the $\chi^2$ for different neutrino masses and then $\Delta \chi^2$ by subtracting its global minimum. Here four different cases are investigated: (a) All the model parameters, including six astrophysical parameters of SN neutrino flux model, the rising time $\tau^{}_{\rm r}$ and the starting time $t^{}_{\rm s}$, are fixed at their true values; (b) The six astrophysical parameters are free, but both $\tau^{}_{\rm r}$ and $t^{}_{\rm s}$ are fixed; (c) Only the rising time is a free parameter; (d) Both the starting time and the rising time are free parameters. Comparing among all the curves in Fig.~2, we can observe that the six astrophysical model parameters have the least impact on the neutrino mass bound. More explicitly, the bound at the $95\%$ CL will change from $m^{}_\nu < 0.807~{\rm eV}$ in the case (a) to $m^{}_\nu < 0.845~{\rm eV}$ in the case (b). The reason can be partially attributed to the fact that these parameters are well determined by the SN neutrino data of higher energies, for which the time-delay effects are rather small. Comparing among the $\Delta \chi^2$ in those four cases, one can clearly see that the neutrino mass bound depends crucially on the starting time when the SN neutrinos are emitted. In the last case, the best-fit value of $m^{}_\nu$ is even located at $0.5~{\rm eV}$, which is mainly due to the statistical fluctuations. The impact of $\tau^{}_{\rm r}$ on the mass bound is visible for small neutrino masses, but becomes insignificant for the large ones. Another important observation is that the distribution of $\Delta \chi^2$ for small neutrino masses is quite flat, indicating the limited capability of the time-delay approach in constraining neutrino masses.
%%%%%%%%%%%%%%%%%%%%%%%%%%%%%%%% Fig. 2 %%%%%%%%%%%%%%%%%%%%%%%%%%%%%%%%%%%%
\begin{figure}
    \centering
    \includegraphics[width=0.6\textwidth]{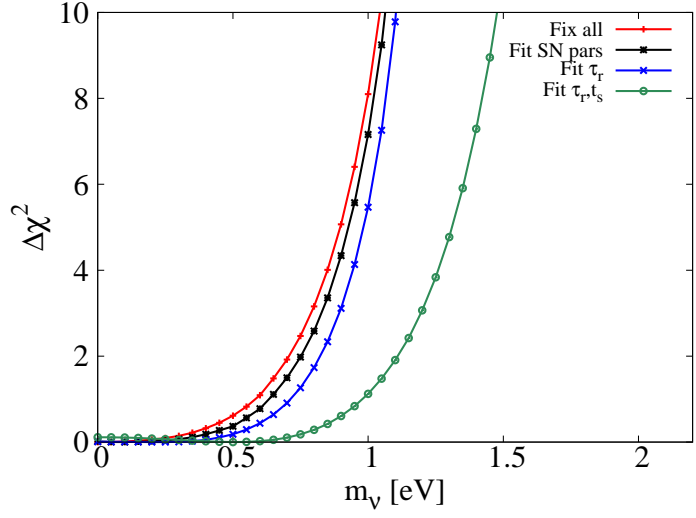}
    \caption{Illustration for the impact of the rising time $\tau^{}_{\rm r}$ and the starting time $t^{}_{\rm s}$ on the neutrino mass bound at JUNO in a single simulation. The artificial data are fitted in four different cases: (a) All the model parameters, including the six astrophysical parameters and two time parameters, are fixed; (b) Only the six astrophysical parameters are free; (c) Only $\tau^{}_{\rm r}$ is free; (d) Both $\tau^{}_{\rm r}$ and $t^{}_{\rm s}$ are taken to be free parameters.}
\end{figure}
%%%%%%%%%%%%%%%%%%%%%%%%%%%%%%%%%%%%%%%%%%%%%%%%%%%%%%%%%%%%%%%%%%%%%%%%%%%%
%%%%%%%%%%%%%%%%%%%%%%%%%%%%%%%% Fig. 3 %%%%%%%%%%%%%%%%%%%%%%%%%%%%%%%%%%%%
\begin{figure}
    \centering
    \includegraphics[width=0.6\textwidth]{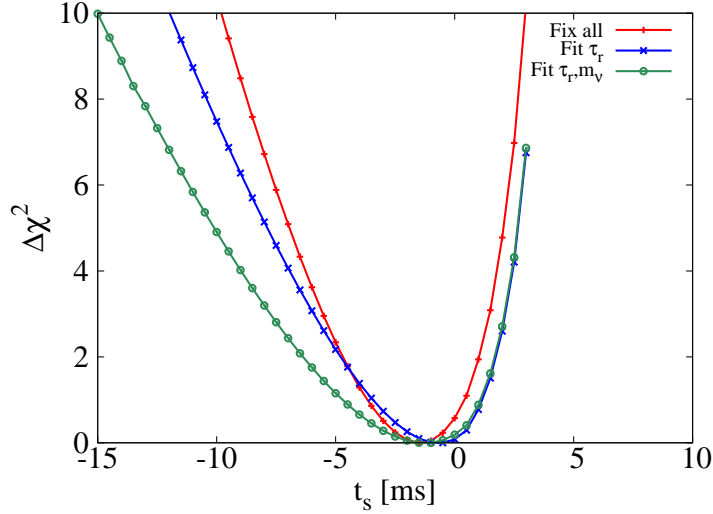}
    \caption{Illustration for the experimental sensitivity to the starting time $t^{}_{\rm s}$ at JUNO. The same set of artificial data as in Fig. 2 has been fitted in three different cases: (a) The neutrino mass $m^{}_\nu$ and the rising time $\tau^{}_{\rm r}$ are fixed at their true values; (b) Only $m^{}_\nu$ is fixed, while $\tau^{}_{\rm r}$ is marginalized over; (c) Both $m^{}_\nu$ and $\tau^{}_{\rm r}$ are taken to be free parameters.}
\end{figure}
%%%%%%%%%%%%%%%%%%%%%%%%%%%%%%%%%%%%%%%%%%%%%%%%%%%%%%%%%%%%%%%%%%%%%%%%%%%%

Based on the same set of artificial data, we also examine the experimental sensitivity to the starting time of neutrino emission at JUNO. In Fig.~3, the data are fitted by $t^{}_{\rm s}$ in three cases, where the neutrino mass $m^{}_\nu$ and the rising time $\tau^{}_{\rm r}$ are fixed or marginalized over. It is worthwhile to note that $t^{}_{\rm s}$ can never exceed the arrival time of the first observed neutrino event, so the curves in Fig.~3 show a common boundary to the right. In the most optimistic situation, the starting time can be determined within a few ${\rm ms}$ at $95\%$ CL, which is comparable to the sensitivity at IceCube in reconstructing the bounce time~\cite{Halzen:2009sm}. Another promising way to determine the absolute emission time of SN neutrinos is to observe the gravitational waves associated with the SN explosion~\cite{Mueller:2003fs,Pagliaroli:2010ik}. If an independent determination of the absolute emission time can be achieved, we may realize the optimistic scenario with fixed $t^{}_{\rm s}$. However, both scenarios of free and fixed time parameters are considered in our studies.

One may also detect the neutronization burst of $\nu^{}_e$ from a core-collapse SN, for which the sharply-rising time can be used to probe the time-delay effects. However, for the current and future huge scintillator detectors, the statistics is limited and the reconstruction of neutrino energies from the elastic neutrino-electron scattering is subject to large uncertainties, because the direction of the final-state electron is not determinable. The charged-current interaction of $\nu^{}_e$ on the carbon target can be implemented as well, but the energy threshold for this process to occur is as high as 17 MeV, leading to a negligible time delay. Another interesting approach to probe absolute neutrino masses based on the time structure of SN neutrinos is to observe the abrupt halt of neutrino signals when a black hole forms during the accretion phase~\cite{Beacom:2000ng,Beacom:2000qy}.

\subsection{Statistical Uncertainties}
%%%%%%%%%%%%%%%%%%%%%%%%%%%%%% Fig. 4 %%%%%%%%%%%%%%%%%%%%%%%%%%%%%%%%%%%%%%
\begin{figure}
  \begin{minipage}[t]{0.5\textwidth}
    \vspace{0pt}
    \centering
    \includegraphics[width=\textwidth]{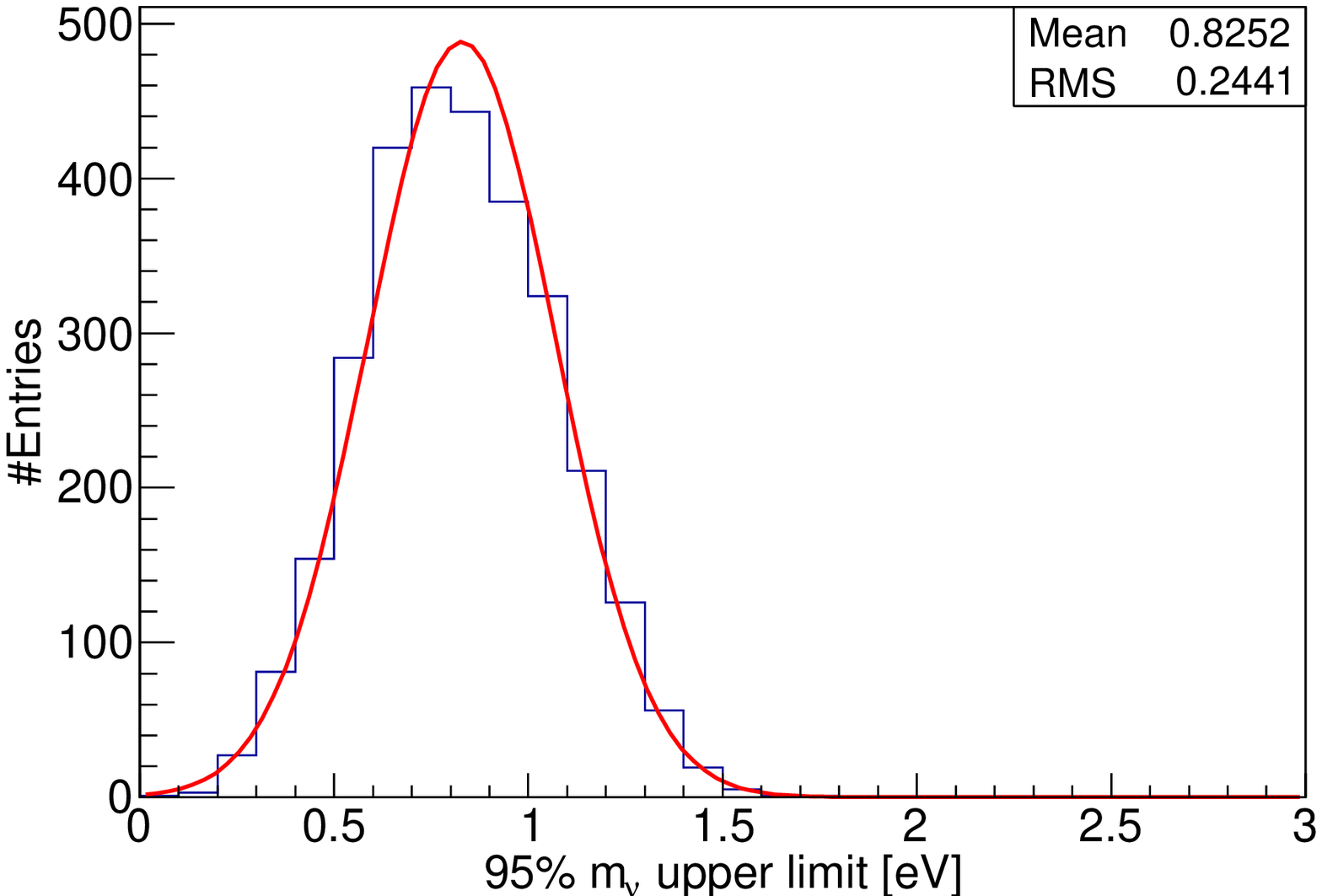}
  \end{minipage}%
  \begin{minipage}[t]{0.5\textwidth}
    \vspace{0pt}
    \centering
    \includegraphics[width=\textwidth]{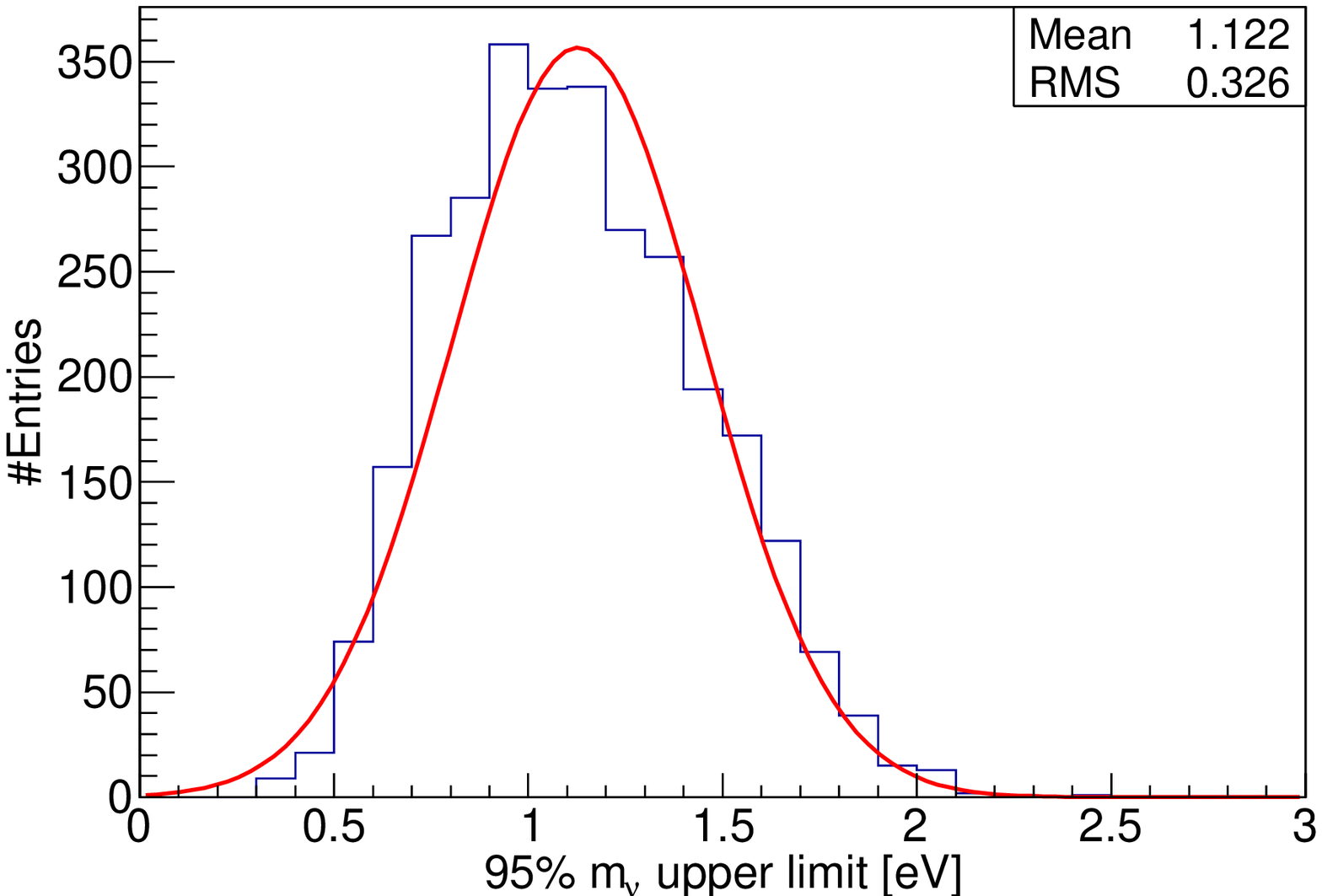}
  \end{minipage}
  \vspace{0.2cm}
\caption{Histogram of the $95\%$ upper limit on neutrino masses for 3000 simulations at JUNO. The model parameters $\tau^{}_{\rm r}$ and $t^{}_{\rm s}$ are fixed for the left panel, while they are free for the right panel. A Gaussian fit to the histogram has been performed and shown together with the mean value and standard deviation.}
\end{figure}
%%%%%%%%%%%%%%%%%%%%%%%%%%%%% Fig. 5 %%%%%%%%%%%%%%%%%%%%%%%%%%%%%%%%%%%%%%
\begin{figure}
  \begin{minipage}[t]{0.5\textwidth}
    \vspace{0pt}
    \centering
    \includegraphics[width=\textwidth]{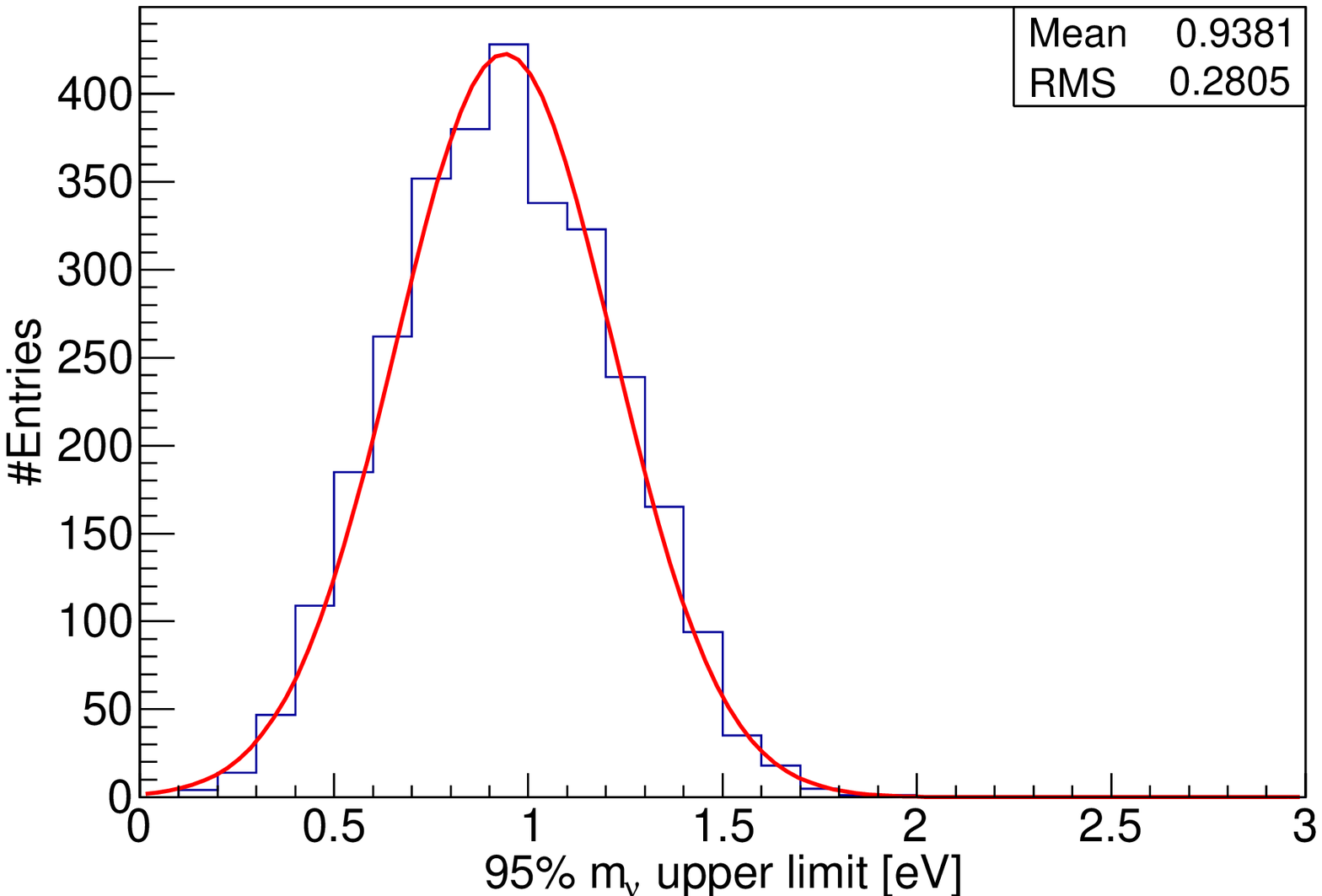}
    \end{minipage}%
  \begin{minipage}[t]{0.5\textwidth}
    \vspace{0pt}
    \centering
    \includegraphics[width=\textwidth]{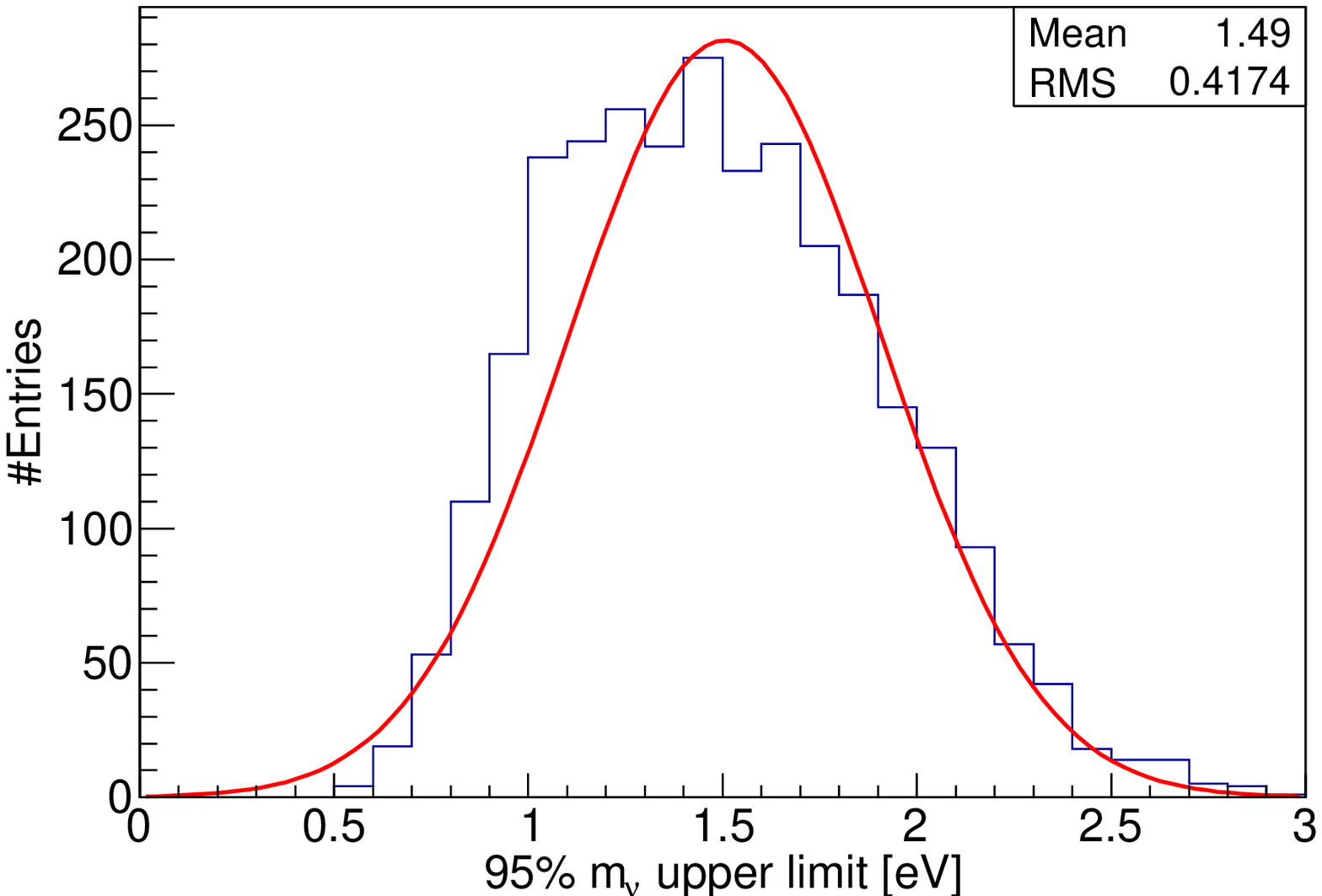}
  \end{minipage}
  \vspace{0.2cm}
\caption{Histogram of the $95\%$ upper limit on neutrino masses for 3000 simulations of the Super-Kamiokande experiment. The model parameters $\tau^{}_{\rm r}$ and $t^{}_{\rm s}$ are fixed for the left panel, while they are free for the right panel. A Gaussian fit to the histogram has been performed and shown together with the mean value and standard deviation.}
\end{figure}
%%%%%%%%%%%%%%%%%%%%%%%%%%%%%%%%%%%%%%%%%%%%%%%%%%%%%%%%%%%%%%%%%%%%%%%%%%%

As we have already mentioned, the neutrino mass bound from one single simulation is not robust, since it suffers from statistical fluctuations. Therefore, we have carried out a large number of simulations for both JUNO and SK, and the final results have been given as histograms in Fig.~4 and Fig.~5, respectively. Furthermore, each histogram is fitted by a Gaussian distribution with the mean value and the standard deviation being shown at the upper-right corner. At the $95\%$ CL, the upper bound on neutrino masses is $m^{}_\nu < (0.83 \pm 0.24)~{\rm eV}$ at JUNO, while $m^{}_\nu < (0.94 \pm 0.28)~{\rm eV}$ at SK. However, if the model parameters $\tau^{}_{\rm r}$ and $t^{}_{\rm s}$ are free parameters, the upper bounds will be relaxed to $m^{}_\nu < (1.12 \pm 0.33)~{\rm eV}$ and $m^{}_\nu < (1.49 \pm 0.42)~{\rm eV}$, respectively.

Since the energy resolution is better and the threshold is much lower at JUNO, compared to those at SK, we expect a more restrictive upper bound. This is really the case, but the improvement is moderate. The main reason is the limited statistics of the low-energy and early-time neutrino events, for which the time-delay effects are significant.

Finally, it is worthwhile to make some remarks on the ultimate sensitivity of future large detectors to the absolute neutrino masses. For this purpose, we have performed the numerical simulations for artificial experiments, which are similar to JUNO but with different energy thresholds, energy resolutions, or target masses.
\begin{itemize}
\item To investigate the impact of energy resolution, we consider two scenarios based on the JUNO setup: (1) no energy smearing; and (2) $\delta E/E = 0.023 + 0.41/\sqrt{{\rm MeV}/E}$, i.e., the same smearing as for SK. The upper bounds on neutrino masses turn out to be $m^{}_\nu < (0.81 \pm 0.24)~{\rm eV}$ and $m^{}_\nu < (0.85 \pm 0.25)~{\rm eV}$ at the $95\%$ CL, respectively. On the other hand, we maintain the energy smearing of JUNO, but assume the energy threshold is $6.5~{\rm MeV}$. In this case, the upper bound $m^{}_\nu < (0.92 \pm 0.27)~{\rm eV}$ becomes much worse than that in the realistic case. Therefore, it is now clear that the energy threshold is the most important reason for the difference between JUNO and SK.

\item Concentrating on JUNO, we now increase its fiducial mass by a factor of two, five and ten, leading to a significant increase in the total number of IBD events. In these three cases, the upper bounds are improved to be $m^{}_\nu < (0.67 \pm 0.20)~{\rm eV}$, $(0.52 \pm 0.15)~{\rm eV}$, and $(0.42 \pm 0.13)~{\rm eV}$, respectively. Note that we have assumed a perfect energy resolution, and that all the model parameters are exactly known. One can observe that increasing the number of neutrino events is the most efficient way to improve neutrino mass bound. Hence, the last bound from ten times larger statistics can be regarded as the ultimate sensitivity of future large detectors.
\end{itemize}
Although the above simulations are not aimed at any realistic detectors, they help us better understand the difference between JUNO and SK, and clarify the limitation of the entire approach to probe absolute neutrino masses.

\subsection{Systematic Uncertainties}
%%%%%%%%%%%%%%%%%%%%%%%%%%%%%%%% Fig. 2 %%%%%%%%%%%%%%%%%%%%%%%%%%%%%%%%%%%%
\begin{figure}
    \centering
    \includegraphics[width=0.7\textwidth]{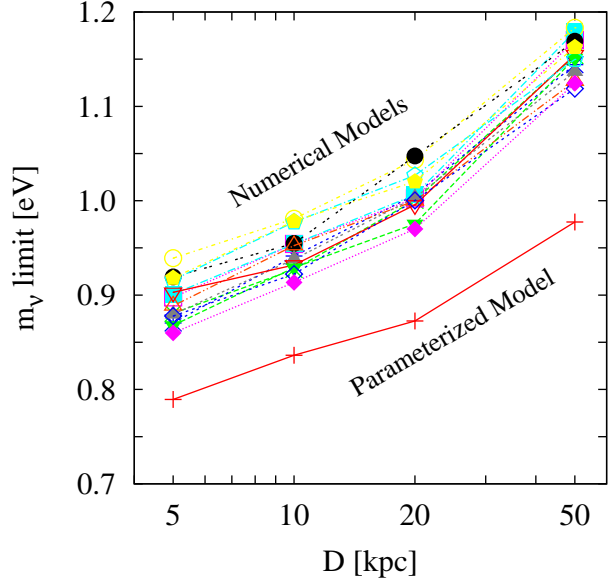}
    \caption{The upper bounds on the absolute scale of neutrino masses at the $95\%$ CL for an SN at distances of $D = 5~{\rm kpc}$, $10~{\rm kpc}$, $20~{\rm kpc}$ and $50~{\rm kpc}$ in the parameterized model from Ref.~\cite{Pagliaroli:2008ur}, and a series of numerical models from Ref.~\cite{Nakazato:2012qf}, in which simulations have been performed for the progenitor-star masses of $M = 13$, $20$, $30$ and $50$ solar masses, the metallicities of $Z = 0.02$ and $0.004$, and the shock revival times of $t^{}_{\rm revive} = 100~{\rm ms}$ and $300~{\rm ms}$. In our calculations, we have chosen fourteen numerical models, since a black hole is formed $842~{\rm ms}$ after bounce in two models with $M = 30$ solar masses and $Z = 0.02$. See Ref.~\cite{Nakazato:2012qf} for more details, and the SN neutrino data are publicly available at http://asphwww.ph.noda.tus.ac.jp/snn/.}
\end{figure}
%%%%%%%%%%%%%%%%%%%%%%%%%%%%%%%%%%%%%%%%%%%%%%%%%%%%%%%%%%%%%%%%%%%%%%%%%%%%
So far, we have concentrated on a very simple description of SN neutrino fluxes in the analytical model~\cite{Pagliaroli:2008ur}. Although such a simplified and parameterized model has intentionally been proposed to capture the main features observed in sophisticated numerical simulations of SN explosions, just a few model parameters are unable to reproduce exactly the fluxes and time structure of SN neutrinos. Therefore, an important issue that one has to address is the systematic uncertainty caused by our ignorance of the true SN model. It is generally difficult to quantitatively describe systematic uncertainties for the lack of knowledge about the SN dynamics, such as the explosion mechanism, but it is practically instructive to look at different SN models and see how the upper bound on neutrino masses changes accordingly.

There are already a few sophisticated simulations of SN neutrino fluxes in the literature~\cite{Mueller:2012is,Nakazato:2012qf}. For illustration, we consider the numerical models from Ref.~\cite{Nakazato:2012qf}, in which a series of SN neutrino light curves and spectra have been calculated by numerical simulations for several progenitor stellar masses ($M = 13, 20, 30$, and $50$ solar masses) and metallicities ($Z = 0.02$ and $0.004$). The simulations have been performed in the spherically symmetric one-dimensional model, and the evolution from the onset of collapse to 20 seconds after the core bounce has been followed by combining the neutrino-radiation hydrodynamic simulations for the early phase and a quasi-static evolutionary calculations of neutrino diffusion for the late cooling phase. The models with a revival time $t^{}_{\rm revive} = 200~{\rm ms}$ have not been considered, since the derived bounds on neutrino masses in these models turn out to be lying between the results for $t^{}_{\rm revive} = 100~{\rm ms}$ and $t^{}_{\rm revive} = 300~{\rm ms}$. In order to make reasonable comparison, the whole duration of neutrino signals in both analytical and numerical models is taken to be 20 seconds. The strategy to calculate the upper bound is the same as before, except that only the neutrino mass is a free parameter in all the models.

In Fig.~6, the upper bounds at the $95\%$ CL in both the parameterized and numerical SN models have been presented for an SN at a typical distance of $D = 10~{\rm kpc}$. One can observe that the bounds in the numerical models span a wide range, which is obviously separated from the bound in the parameterized model. The reason why the upper bound in the parameterized model is the best can be understood by having a close look at the number of events in the early-time and low-energy regions. For one single simulation, we have found $57$ events before $t = 0.1~{\rm s}$ and below $E^{}_\nu = 10~{\rm MeV}$ in the parameterized model, while about 20 events in the numerical models. In the numerical models, neutrino energy spectra in the early time significantly deviate from the thermal one and take on long high-energy tails~\cite{Nakazato:2012qf}, so we have more high-energy neutrino events, for which the time-delay effects are small. The systematic uncertainty caused by SN models from our calculations could be as large as $0.2~{\rm eV}$, which is comparable to the statistical error. But it is worthwhile to stress that more numerical models should be considered to make the systematic error more reliable.

To see how the result will be modified for different SN distances, we also compute the upper bounds for $D = 5~{\rm kpc}$, $20~{\rm kpc}$ and $50~{\rm kpc}$ for illustration. The last one represents the distance to SN 1987A in the Large Magellanic Cloud. Now it is straightforward to see how the mass bound becomes worse if the SN is located at a larger distance. For an SN in the Large Magellanic Cloud, the bound is found to be $m^{}_\nu < 0.98~{\rm eV}$ in the parameterized model, and $m^{}_\nu < (1.1 \cdots 1.2)~{\rm eV}$ in the numerical models. Generally speaking, the upper bound can be improved for a closer SN or worsened for a farther one, just by scaling up or down the number of neutrino events. But it should be noticed that a longer distance implies a more significant time delay, which will compensate somehow the reduced number of neutrino events.

Before finishing this section, we would like to make some comments on the detector-related systematics. The non-SN backgrounds for SN IBD events are dominated by the antineutrinos from nearby nuclear reactors, which are estimated to be 80 events per day for JUNO without efficiency cuts, and therefore are negligible for SN neutrinos with several thousands of events in ten seconds. On the other hand, the SN-related backgrounds are from the accidental coincidence of the neutrino-electron or neutrino-proton scattering singles. Considering the time and space correlation for the real IBD events, we can also safely neglect the SN-related backgrounds for SN IBD event studies. Finally, the IBD selection efficiency can be as large as 99\%, and the accuracy of the energy scale will be better than 1\%. Therefore, in our current study, we neglect all the detector-related systematics for JUNO. However, in order to accurately address this problem, we have to carry out realistic simulations for SN neutrinos in the JUNO detector, which are now under consideration and will be presented separately in future works.

As for SN neutrino detection at SK without Gd-doping~\cite{Ikeda:2007sa}, the techniques for vertex and energy reconstruction of the IBD signals are the same as those used in the analysis of solar neutrinos in Ref.~\cite{Hosaka:2005um}, where one can find that the final selection efficiency reaches $40\%$ at the energy threshold $6.5~{\rm MeV}$ and increases to $62\%$ for the energies above $12.5~{\rm MeV}$. Since there is no neutron tagging, the non-IBD SN events contribute to the main background, which can be estimated as $9\%$ from the neutrino-oxygen interactions (including both neutral- and charged-current channels) and $3\%$ from the elastic neutrino-electron scattering~\cite{Ikeda:2007sa}. Therefore, the detector-related systematic uncertainties at SK are significant and should be included in a dedicated study. 

Two methods for neutron tagging in SK have been proposed in Ref.~\cite{Watanabe:2008ru}. The first one is to implement neutron captures on gadolinium, which yield 8 MeV gamma rays. If 2.4 liters of $0.2\%$ ${\rm Gd}{\rm Cl}_3$ water-solution are added into the SK detector, a neutron-tagging efficiency of $66.7\%$ can be achieved for the events above $3~{\rm MeV}$ and the corresponding background can be reduced to $2\times 10^{-4}$. The second one is to consider neutron captures on hydrogen, which yield 2.2 MeV gamma rays. A forced trigger system is introduced to take 500 $\mu$s of data with no threshold requirement immediately after any primary events, such that the 2.2 MeV $\gamma$'s can be statistically identified as correlated in time with an energetic primary event. In assumption of a uniform distribution of the 2.2 MeV $\gamma$'s in the SK detector, the neutron efficiency is found to be approximately $20\%$, and the background reduction in this approach is at the level of $3 \times 10^{-2}$. 

In summary, the backgrounds and detector-related systematics for SK are important and should be taken seriously, while those for JUNO can be safely neglected. We expect that the situation for SK can be greatly improved in the future with an efficient neutron tagging.

\section{Summary}

Recent years have seen great progress in experimental neutrino physics. In particular, the smallest neutrino mixing angle has been precisely measured in the Daya Bay reactor neutrino experiment. The primary goals of future neutrino oscillation experiments will be the determination of neutrino mass ordering and the discovery of leptonic CP violation. Unfortunately, the oscillation experiments are insensitive to the absolute neutrino masses. As is well known, the tritium beta decays and neutrinoless double-beta decays could provide us with useful information about neutrino masses in the sub-eV region. Currently the tightest bound on the sum of neutrino masses $\Sigma < 0.23~{\rm eV}$ comes from the cosmological observations, which however may suffer from large systematic uncertainties. Therefore, an independent way to probe neutrino masses at the sub-eV scale is desirable and important.

In the present work, we consider the possibility to constrain neutrino masses by observing galactic SN neutrinos at the JUNO detector. As the largest scintillator detector over the world, JUNO will register about five thousand neutrino events in the inverse-beta decay channel for a galactic core-collapse SN at a distance of $10~{\rm kpc}$. Since the arrival time and neutrino energy can be well measured at JUNO, the distortion in the time distribution of SN neutrino events caused by the delay of flight time is sensitive to the absolute scale of neutrino masses. Based on a simple but useful model of SN neutrino fluxes and the maximum likelihood analysis, we have carried out a number of simulations to explore the upper bound on absolute neutrino masses at JUNO. In assumption of a nearly-degenerate mass spectrum and a normal mass ordering, it is found that $m^{}_\nu < (0.83 \pm 0.24)~{\rm eV}$ at the 95\% CL can be reached, where the mean value and standard deviation are shown to account for the statistical fluctuation. For comparison, we find that the bound in the Super-Kamiokande experiment is $m^{}_\nu < (0.94 \pm 0.28)~{\rm eV}$ at the same CL. Different from previous works, the impact of astrophysical model parameters on the neutrino mass bound has been emphasized and studied in more detail. Moreover, the statistical uncertainties of the mass bound have also been taken into account. The systematic uncertainties induced by the model dependence are illustrated by calculating the mass limits in a series of numerical models of SN neutrinos.

Although the neutrino signals from a galactic core-collapse SN explosion depend very much on the intrinsic properties of the progenitor star, such as the distance and the initial mass, the rapidly rising or falling feature in the time structure of SN neutrinos can be used to extract useful information about neutrino masses. For instance, the early-time neutrino burst from neutronization and the abrupt halt of neutrino signals due to the black hole formation will be advantageous for this purpose. We hope to return to those possibilities in the near future.

\section*{Acknowledgements}

This work was supported in part by the National Natural Science Foundation of China under Grant Nos. 11135009 and 11305193, the Strategic Priority Research Program of the Chinese Academy of Sciences under Grant No. XDA10010100, the Innovation Program of the Institute of High Energy Physics under Grant No. Y4515570U1 and the CAS Center for
Excellence in Particle Physics (CCEPP).

\end{document}